\documentclass[draft]{agujournal2018}
\usepackage{apacite}
\usepackage{url} 

\draftfalse

\journalname{Geophysical Research Letters}

\begin{document}

%
%

\title{The fine structure of the subsolar MPB current layer from MAVEN observations: Implications for the Lorentz force}




\authors{G. Boscoboinik\affil{1}, C. Bertucci\affil{1,3}, D. Gomez\affil{1,3}, L. Morales\affil{2,3}, C. Mazelle\affil{4}, J. Halekas\affil{5}, J. Gruesbeck\affil{6}, D. Mitchell\affil{7}, B. Jakosky\affil{8}, E. Penou \affil{4}}

\affiliation{1}{IAFE, UBA CONICET, Buenos Aires, Argentina}
\affiliation{2}{INFIP, UBA CONICET, Buenos Aires, Argentina}
\affiliation{3}{Departament of Physics, FCEyN, UBA, Buenos Aires, Argentina}
\affiliation{4}{IRAP, UPS CNRS CNES, Toulouse, France}
\affiliation{5}{University of Iowa, Iowa City, IA, USA}
\affiliation{6}{GSFC, Greenbelt, MD, USA}
\affiliation{7}{SSL, University of California, Berkeley, USA}
\affiliation{8}{LASP, University of Colorado, Boulder, CO, USA}

\correspondingauthor{Gabriela Boscoboinik}{gboscoboinik@iafe.uba.ar}




\begin{keypoints}
\item We analyse the fine structure of the current layer at the Martian Magnetic Pileup Boundary (MPB) in the subsolar sector.

\item MPB thickness is of the order of the solar wind proton inertial length or convective Larmor radius in the magnetosheath. 

\item The work done by the Lorentz Force suggests that solar wind ions can be stopped by magnetic pressure at the MPB.

\end{keypoints}

%
%




\begin{abstract}
We report on the local structure of the Martian subsolar Magnetic Pileup Boundary (MPB) from minimum variance analysis of the magnetic field measured by the Mars Atmosphere and Volatile EvolutioN (MAVEN) spacecraft for six orbits. In particular, we detect a well defined current layer within the MPB’s fine structure and provide a local estimate of its current density which results in a sunward Lorentz force. This force accounts for the deflection of the solar wind ions and the acceleration of electrons which carry the interplanetary magnetic field through the MPB into the Magnetic Pileup Region. We find that the thickness of the MPB current layer is of the order of both the upstream (magnetosheath) solar wind proton inertial length and convective gyroradius. This study provides a high resolution view of one of the components of the current system around Mars reported in recent works.

\end{abstract}

\section*{Plain Language Summary}
We investigate the fine structure of the current layer associated with the outer boundary of the Martian induced magnetosphere in the subsolar sector from selected MAVEN magnetic field and solar wind plasma observations. We measure the variance of the magnetic field across the boundary to detect the current layer and measure the strength of the current that circulates there. The current density we obtain is such that its derived Lorentz force is strong enough to stop the solar wind ions at the outer boundary of Mars magnetosphere. On the other hand, this force would push the solar wind electrons and the interplanetary magnetic field frozen into the electron plasma into the induced magnetosphere.
We also find that the thickness of this current layer in terms of typical lengths of the solar wind ion plasma is similar to the thickness of the terrestrial magnetopause.

\section{Introduction}

Mars (1$R_M$ = 3390 km) has either no or negligible present global magnetic field ($ | M | <2 \times 10 ^{21} \textrm{G} \cdot \textrm{cm }^3 $) \citep{acuna}. This makes the solar wind interact directly with its ionosphere and the charged particles from its exosphere. The Martian atmosphere is mostly composed of carbon dioxide ($ \textrm{CO}_2$), carbon monoxide (CO), argon (Ar) and molecular nitrogen ($ \textrm{N}_2$). In smaller quantities are found molecular and atomic oxygen ($ \textrm{O}_2$ and O), nitrogen monoxide (NO), atomic nitrogen (N) and helium (He). The relative proportions of the species that populate the atmosphere vary with altitude. In particular, in the high atmosphere (altitudes greater than 200 km), which is the region of interest for this work, the dominant species are atomic oxygen, molecular oxygen and hydrogen \cite{anderson1971, anderson1974, mahaffy}.

The interaction of the solar wind with Mars' atmosphere produces the so-called \textit{induced magnetosphere}, a region where the solar wind flow and field are disturbed by the presence of the planet.
With an areocentric distance of approximately $2~R_M$ for the bow shock (BS) and $1.3~R_M$ for the MPB (between 800 and 1000 km), the magnetosphere of Mars is one of the smallest of the solar system \citep{moses1988}. 
However, it is in this small portion of space that most of the solar wind's energy and momentum are transferred to the planetary plasma.
Recent estimates of atmospheric escape on Mars \citep{jakosky_MAVEN} suggest that the interaction with the solar wind has played a significant role in the removal of water from Mars for billions of years. In this context, the study of these electric fields is essential to understanding the processes of energy and momentum transfer from the solar wind to the plasma of planetary origin that lead to atmospheric escape.

The supermagnetosonic nature of the solar wind needs a bow shock to form ahead of the obstacle to avoid it. Downstream from the BS, the solar wind plasma is mostly subsonic and significantly hotter. Also in this region -named magnetosheath- the magnetic field is highly variable due to the presence of turbulence \citep{Ruhunusiri2017} and waves generated from electron and ion instabilities taking place both upstream and locally. 
In the lower part of the magnetosheath, the solar plasma slows down further as it increasingly incorporates cold protons and heavier ions from Mars’ exosphere. 
These particles, being relatively slow, heavy and numerous compared to the solar wind, decrease the average speed of the solar wind \citep{szego} in areas where the influence of crustal magnetic fields is negligible \citep{connerney2001}. 
This deceleration precedes a change in the composition of the plasma, from solar wind ions to heavy ions of planetary origin, at the Ion Composition Boundary (ICB), which on the dayside is almost coincident with the MPB. \cite{Breus1991,Sauer1994,matsunaga2017,halekas2018-ICB,Holmberg2019}

In areas where crustal magnetic field can be ignored, the mass-loading causes the frozen in interplanetary magnetic field to increase in the subsolar region and to drape around the planet. On the dayside, the increase in the magnetic field strength has been found to be a permanent feature although single spacecraft magnetic field time series suggest a variety of values for this gradient. Following the nomenclature of a similar structure at active comets \citep{Neubauer1987} the layer where the magnetic field strength gradient occurs received the name Magnetic Pileup Boundary \citep{acuna}. Pre MAVEN measurements \cite{dubinin2008, bertucci2011} have shown that the MPB is located between the region dominated by the solar plasma -the magnetosheath- from that governed by the plasma of planetary origin -the Magnetic Pileup Region (MPR), also called Induced Magnetosphere-, which is characterized by a strong and organized magnetic field of solar origin as a result of pileup and draping \citep{bertucci2003a}. Once again these features apply for regions where crustal fields are not important.  
The MPR lies above the ionospheric boundary, sometimes called ionopause, its lower limit. Below the ionopause, the frequency of collisions between particles increases above the typical frequencies of a plasma, allowing the diffusion of the magnetic field.

In addition to the former, other features allow the detection of the MPB at Mars and other atmospheric bodies \citep{bertucci2011}: a marked increase in the magnitude of the magnetic field (by a factor of 2 or 3) followed by a decrease in the magnetic field fluctuations, a decrease in the temperature, velocity and density of the solar wind ions and suprathermal electrons and an increase in the total plasma density as an increase in the number of charged particles of planetary origin. 
These features have allowed for statistical studies on its average location and shape \cite{Vignes, trotignon2006, edberg2008}. 
So far the fine structure of the MPB has been studied from single spacecraft observations \citep{bertucci2005} or multifluid simulations of high spatial resolution \citep{Harnett2007}.
\citet{bertucci2005} applied minimum variance analysis (MVA) \citep{sonnerup_scheible} to MGS magnetometer observations near the terminator and found that inside the MPB there is a layer of typically 100 km where the magnetic field vector rotates on a plane that is nearly perpendicular to the boundary normal obtained from the MPB static fit. The surface and volume current densities were 6.5 $\times 10^6$ nA/m and 81 nA/m$^2$ respectively, comparable to values obtained from multi fluid simulations.
Unfortunately, this work was limited to high solar zenith angles or SZA (i.e., larger than 30º) because of the geometry of MGS pre-mapping orbits. But also, the lack of ion measurements precluded any local estimate of relevant plasma length scales necessary to assess the origin of the detected currents.

In previous studies it has been shown that in the different regions of the Martian magnetosphere different terms of the electric field prevail \citep{dubinin2011}. With the arrival of the Mars Atmosphere and Volatile Evolution (MAVEN) mission, reliable, high resolution particle and magnetic field measurements have become available for a deeper analysis of the macroscopic current systems within Mars' magnetosphere. 
\citet{halekas_forces} obtained averaged values of the current density and the derived Lorentz force ($\mathbf{J}\times \mathbf{B}$) around the MPB by estimating the curl of the magnetic field accumulated in static bins with a resolution of 500 km in the \textit{x-y} plane and 2000 km in \textit{z}. 

More recently \citet{Ramstad2020} have reported on a global, coupled current system at Mars by computing $\mathbf{J} = \frac{1}{\mu_0} (\nabla \times \mathbf{B})$ as center-point differences for every location of two 3D magnetic field map. This map was obtained averaging the magnetic field obtained over 9814 orbits with a grid spacing of 0.1 $R_M$ or 0.2 $R_M$ depending on the altitude. 

As we have access to high resolution data we can determine more precisely where the current sheet is located inside the MPB and obtain its thickness in order to understand where this current is originated.

In the absence of collisions, local particle acceleration is produced by electric fields. Within the framework of a multifluid plasma, the equation of motion for each species $s$ is

\begin{equation}
m_s n_s \frac{d \textbf{v}_s}{dt} =  q_s n_s (\textbf{E} + \textbf{v}_s \times \textbf{B}) - \nabla \cdot \mathbf{P}
\label{eq:momentum_s}
\end{equation}
where $m_s$, $q_s$ are the individual mass and electric charge of particles of species $s$, $n_s$ and $\textbf{v}_s$ are the particle density and velocity of the fluid formed of $s$-particles and \textbf{P} is the pressure tensor. If we assume a plasma made of electrons and a single ion species, quasi-neutrality dictates that $n_e = n_i = n$ and the current density is simply given by $\textbf{j} = e n (\textbf{v} - \textbf{v}_e)$. If we further assume that the electron mass is negligibly small ($m_e\approx 0$), the equation for electrons reduces to a force equilibrium given by 

\begin{equation}
    \textbf{E} = - \textbf{v}\times \textbf{B} + \frac{1}{en} (\textbf{j}\times \textbf{B} - \nabla \cdot \mathbf{P}_e)
    \label{eq:ohm}
\end{equation}
This equation is also known as the generalized Ohm's law.

The bulk velocity of the plasma is $\textbf{v} = \textbf{v}_i$, since momentum is fully carried by ions in this approximation. The equation of motion for the ions, after replacing Eqn~\ref{eq:ohm} into Eqn~\ref{eq:momentum_s} and using the identity $\textbf{j} \times \textbf{B} = \frac{1}{\mu_0} (\textbf{B} \cdot \nabla) \textbf{B} - \nabla\frac{B^2}{2\mu_0} $, reduces to

\begin{equation}
m_i n \frac{d \textbf{v}}{dt} =  \frac{1}{\mu_0} (\textbf{B} \cdot \nabla) \textbf{B} - \nabla\frac{B^2}{2\mu_0} - \nabla \cdot (\mathbf{P}_e + \mathbf{P}_i)
\label{eq:momentum_i}
\end{equation}
where the first term on the RHS is the magnetic tension force, the second term is the magnetic pressure force and the last term is the total thermodynamic pressure. The magnetic pressure is directly proportional to the square of the magnetic field and inversely proportional to the thickness of the MPB. In contrast, the term of the tension, while also directly proportional to the square of the magnetic field, is inversely proportional to the curvature radius of the magnetic field lines.

In the present work we analyse MAVEN data to identify and characterize the local structure of the Martian subsolar MPB. Then we apply MVA to MAVEN magnetic field measurements to estimate the local current density flowing along the MPB and its associated Lorentz force in order to evaluate its importance in the plasma dynamics around the boundary. In section \ref{section:M&D} we describe the data and methods used, the results are displayed in section \ref{section:results} and are discussed in section \ref{section:discussion}.

\section{Methods and Data}
\label{section:M&D}

We analysed six subsolar MPB crossings between October 2015 and November 2017 from MAVEN data. The magnetic field data measured by the Magnetometer (MAG) \citep{MAG} has a 32Hz sampling rate. The solar wind electron data from the Solar Wind Electron Analyzer (SWEA) \citep{SWEA} measures electrons in an energy range between 3 eV and 4600 eV with a 2 s resolution. The Solar Wind Ion Analyzer (SWIA) \citep{SWIA} provided the solar wind proton data in an energy range between 25 eV and 25 keV with a 4 s resolution. 

\subsection{Methodology}

The Minimum Variance Analysis, or MVA for a single spacecraft \citep{sonnerup_scheible} is a technique widely used to find the normal vector for a one-dimensional discontinuity from magnetic field measurements obtained by the probe across the boundary (e.g. \citet{knetter}).
The main purpose of the MVA is to estimate the normal to a one-dimensional current sheet in a collisionless plasma. This is achieved by determining the eigenvectors and eigenvalues of the covariance matrix defined as $M^B_{\mu\nu} \equiv \langle B_\mu B_\nu \rangle - \langle B_\mu \rangle \langle B_\nu\rangle $
in terms of the magnetic field data and the coordinate system in which the data is presented, then find its three eigenvalues $ \lambda_i $ and their corresponding eigenvectors $ \textbf{x} _i $. The eigenvector corresponding to the smaller eigenvalue ($ \textbf{x} _3 $ and $ \lambda_3 $), is the estimate for the direction of the normal vector to the current sheet and $ \lambda_3 $ represents the variance of the magnetic field component in that direction.
In general, for any set of vectors $ \{\textbf{B} ^{(m)}\} $ across a transition layer, the set of $ M^B_ {\mu \nu}$ eigenvectors provides a convenient coordinate system to analyse the data. It must be noted that the variance matrix $ M ^{B}_{ \mu \nu}$ is independent of the temporal order of the measured vectors.

In the present work MVA is applied to the MAG data in the MPB in order to obtain an estimate of the normal vector to this boundary and therefore to the associated current sheet.

Another estimate of the normal vector to the MPB can be obtained from the conic section fit representing its average position (e.g \cite{Vignes}). The functional form of the fit is the following:
\begin{equation}
    r = \frac{L}{1+\epsilon \cos(\theta)}
    \label{eq:vignes}
\end{equation}
where $r$ and $\theta$ are polar coordinates with origin at $x_0$, $\epsilon$ is the eccentricity and \textit{L} is the semi-latus rectum. From this fit it can be therefore obtained the normal vector to the surface $\hat n$.

In this first study we deliberately selected crossings that show an apparently sharp increase in the magnetic field amplitude, are located on the northern hemisphere and all have Solar Zenith Angle (SZA) $<$ 30º. The crustal magnetic field according to the model by \citet{Cain2003} does not exceed 10$\%$ of the total field in the analysed crossings.
These crossings occurred in the span of more than one Martian year and have varied solar wind conditions and heliocentric distance.

\subsection{MPB Identification in a Case Study}

\begin{figure}[h]
    \centering
    \includegraphics[width=30pc]{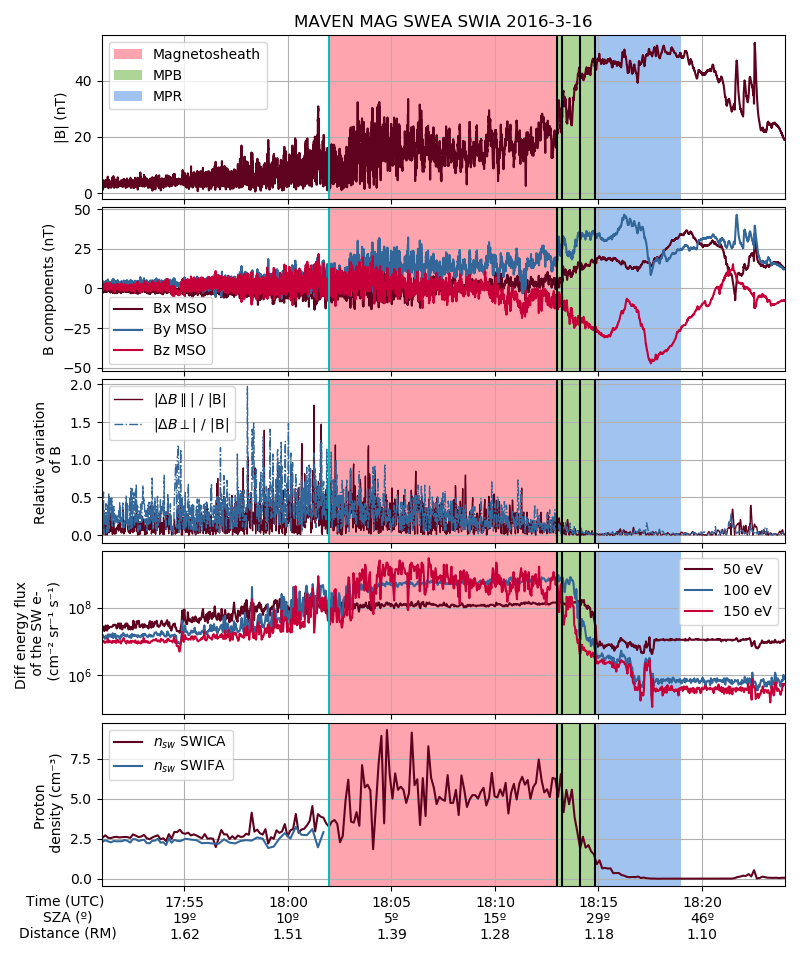}
    \caption{Time series of the magnetic field and plasma data from MAVEN for the March 16th 2016 crossing. From top to bottom: Magnetic field magnitude, Magnetic field components, Relative variation of the magnetic field, Differential energy fluxes for solar wind electrons and Solar wind proton density. The MPB is shaded in green.}
    \label{fig:mpb}
\end{figure}

Fig. \ref{fig:mpb} shows a time series of magnetic field and plasma data from MAVEN near the MPB for one of the orbits analysed in this work. All vector magnitudes are represented in the Mars-centered Solar Orbital (MSO) coordinate system, in which the $\hat x$ axis points from Mars towards the Sun, the $\hat y$ axis points antiparallel to Mars' orbital velocity and $\hat z$ completes the right-handed coordinate system. 

Between 17:40 and 18:40 UTC on March 16th, 2016 MAVEN headed from the undisturbed solar wind to Mars, crossing the bow shock a few minutes after 17:50 UTC and the MPB near the subsolar point around 18:10 UTC. Then, MAVEN continued within the induced magnetosphere and ionosphere and at 18:30 UTC entered the region of the magnetic tail.

For the identification of the MPB we rely on the criteria described by \citet{bertucci2011}:
a sharp increase in the magnetic field strength by a factor of 2-3, a sharp decrease in the magnetic field fluctuations, a sharp enhancement of the magnetic field draping, a decrease in the temperature of electrons and a decrease in the solar wind proton density.

In order to determine the MPB thickness, we selected four times which we called $t_1$, $t_2$, $t_3$, $t_4$ so that outside the interval between $t_1$ and $t_4$ MAVEN would be unambiguously outside the MPB while in the interval between $t_2$ and $t_3$ MAVEN would be inside the MPB. In this interval we observe the defining characteristics of this boundary. 
The times thus determined were $ t_1 = $ 18:13:00 UTC, corresponding to an altitude of 734 km and an SZA of 23º, $ t_2 = $ 18:13:13 UTC with altitude 720 km and SZA 24º, $ t_3 = $ 18:14:06 UTC with altitude 663 km and SZA 26º and $ t_4 = $ 18:14:51 UTC with altitude 615 km and SZA 29º.

In this interval we observe the drastic changes in the plasma near the MPB: the magnetic field changes direction while its magnitude goes from 20 nT to 45 nT in less than 2 minutes. We also observe that the relative variations of $\mathbf{B}$ (both parallel and perpendicular to the mean field) cease abruptly. This decrease is due to the diminishing amplitude of the fluctuations as well as the increase in magnetic field magnitude.
The differential energy fluxes decrease in a range from one to two orders of magnitude in the MPB depending on the electron energy, which is consistent with the electron impact ionization described by \citet{crider2000}. 
The solar wind proton density decreases from 6 cm$^{-3}$ down to the instrumental noise for protons with energies above 25 eV.

\section{Results}
\label{section:results}
\begin{figure}[h]
    \centering
    \includegraphics[width=30pc]{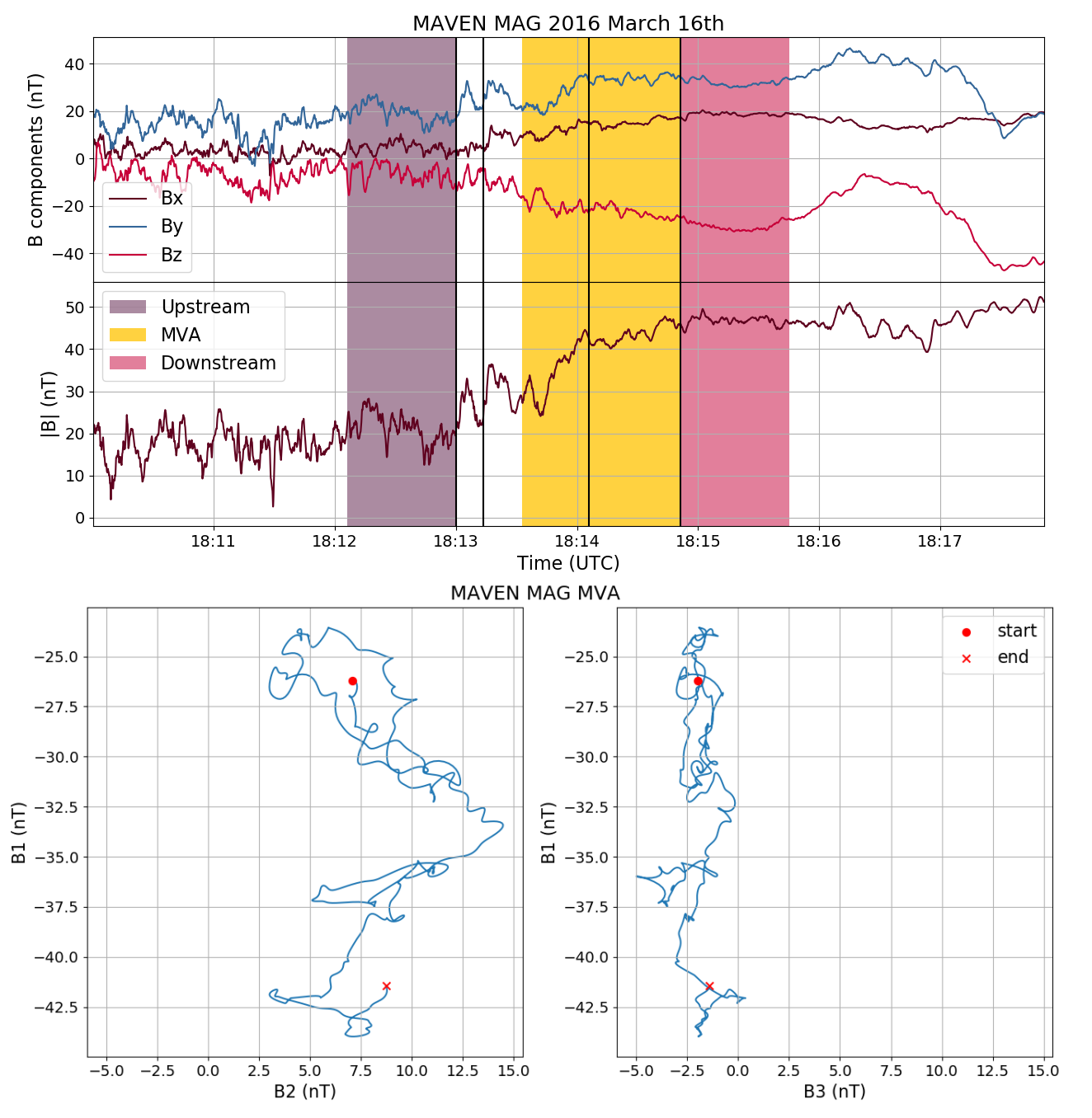} 
    \caption{Magnetic field components in MVA coordinates and amplitude (up). The upstream and downstream intervals from the MPB and interval where MVA was applied are shaded. Magnetic hodograms (N=922, $\lambda_2/\lambda_3$ = 9.8) depicting the magnetic field projection on the planes ($ \hat{e} _1 $, $ \hat e_2 $) and ($ \hat e_1 $, $ \hat e_3 $) in the interval where MVA was applied (down). The start point is marked with a circle and the end point with a cross.}
    \label{fig:zoom}
\end{figure}

Once the times $ t_1 $, $ t_2 $, $ t_3 $ and $ t_4 $ delimiting the MPB were identified we applied MVA in the interval 18:13:37 - 18:14:06 UTC (shown shaded in Fig. \ref{fig:zoom}); the data consisted of 922 high resolution measurements. 
We chose this interval in order to have the best MVA result within the MPB thus identified.
Looking at the upper panel of the Fig. \ref{fig:zoom}, where the magnetic field components are plotted, we can anticipate that the minimum variance direction will be approximately parallel to the $ \hat x $ axis. We also see that the field points mainly in the $ \hat y $ and $ \hat z $ directions, so we can anticipate that \textbf{B} in the MPR will be mostly tangential.

The intermediate-to-minimum eigenvalue ratio for the analysed crossing is $ \lambda_2 / \lambda_3 = 9.48 $, which ensures that the minimum variance vector is well defined \citep{knetter}.

The normal obtained with this method is $ \mathbf{x} _3 = \hat n _\textrm{MVA} = (0.920,	-0.302,	0.251) $ with angular error 0.65º, 
that is, differing by 23º from the $ \hat x $ axis. The mean magnetic field component along the normal is $ \langle B_n \rangle = -2.06 \pm 0.08 ~\textrm{nT} $ and the mean magnetic field magnitude is $B_0 = |\langle \mathbf{B} \rangle| = 34.79$, the quotient between both being $\langle B_n\rangle/B_0 = 0.06$ which is consistent with our assumption that the magnetic field in the MPR would be nearly tangential. 
The angle between the mean magnetic field vector $ \langle \mathbf{B} \rangle $ and the normal is $ \theta_B = 93$º, that is, the magnetic field is almost tangential and lies mostly in the ($ \hat{e} _1 $, $ \hat{e} _2 $) plane. 
The hodograms in Fig. \ref{fig:zoom} show the magnetic field projection on the planes ($ \hat{e} _1 $, $ \hat e_2 $) and ($ \hat e_1 $, $ \hat e_3 $) in the interval where MVA was applied (between 18:13:37 and 18:14:06 UTC). The hodogram to the right (depicting the projection $ \hat e_1 $, $ \hat e_3 $) has an elongated shape, consistent with a good eigenvalue ratio and the plane containing the normal being well defined.

In order to obtain the MPB thickness, we calculated the angle $ \theta_v $ between the average spacecraft velocity $\langle v_\textrm{sc}\rangle$ in the MPB and the normal; the calculation yielded $ \theta_v = 117 $º. This means that MAVEN motion was almost parallel to the MPB.

Once we have the normal we can estimate the thickness of the MPB \textit{h} assuming that the boundary is one dimensional and static. 
We then approximate $ h = | (\mathbf{r} _ {in} - \mathbf{r} _ {out}) \cdot \hat{n} | $, where $ \mathbf{r} _ {in} $ is the position of the spacecraft when entering the MPB and $ \mathbf{r} _ {out} $ is the position when leaving; being that it is not uniquely defined, we actually approximate a minimum thickness corresponding to the interval between $ t_2 - t_3 $ and a maximum thickness in the interval $ t_1 - t_4 $. In this way, we obtained $ h_{23} = 82 ~\textrm{km} $ and $ h_{14} = 174 ~\textrm{km} $. These values are comparable to both the magnetosheath solar wind proton inertial length ($ \lambda = c / \omega_{pi} = 97.9 ~\textrm{km} $) and the magnetosheath convective proton gyroradius ($r_g = \frac{m v_\perp}{|q| B} = 68.4 km$). 
The solar wind proton inertial length was calculated from SWIA data, as $ \omega_{pi} $ is the proton plasma frequency obtained using the mean proton density in the upstream region (shown shaded in Fig. \ref{fig:zoom}). On the other hand, for obtaining the magnetosheath convective proton gyroradius we considered $B$ as the average magnetic field and $v_\perp$ as the velocity perpendicular to $B$ in the upstream region. 

For the other MPB normal estimate, fitting the mean MPB position with an ellipsoid given by equation \ref{eq:vignes}, we used the parameters $ x_0 = 0.78 R_M $ and $ \epsilon = 0.9 $ given by \citet{Vignes}. Requiring that the ellipsoid contains the point through which the spacecraft passes at \textit{t} = 18:13:49, the semi-latus rectum is $ L = 0.87 R_M $. We chose this point as it corresponds to half the interval which delimits the current sheet.

The normal thus obtained is $\hat{n}_\textrm{fit}$ = (0.856, -0.066, 0.512), a value that differs by 21º from that of the normal obtained by applying MVA and by 31º from the $ \hat x $ axis. 
The mean value of the magnetic field along this normal is $ \langle B_3 \rangle = -1.72 ~\textrm{nT} $, which when comparing it to $B_0$ yields $ \langle B_3 \rangle / B_0 = 0.05$. 
The angle $ \theta_B $ between the mean magnetic field vector and this normal is $ \theta_B$ = 92.8º. We observe again that the magnetic field is almost tangential to the boundary.

We obtained $ \theta_v = 101 $º, which is consistent with the idea that the motion of the spacecraft is almost parallel to the surface of the MPB. 

In the same way as before, we estimated the MPB thickness \textit{h} in the intervals $ t_2 - t_3 $ and $ t_1 - t_4 $. In this case, the obtained values are smaller, which is to be expected given that the angle $ \theta_v $ is smaller, yielding $ h_{23} = 34$ km and $ h_{14} = 71$ km.

In general, we consider the results derived from MVA to be more representative of reality, since this method is based on the local properties of the magnetic field at the time of the crossing. Nonetheless, results show a good agreement between the local (MVA) and the macroscopic (fit) normals and in both cases we observe that the normal points mostly along $ + \hat x $, which is consistent with a SZA close to 25º.

In table \ref{t:thickness} are displayed the thickness of the MPB obtained from MVA, the solar wind convective proton inertial length and the convective Larmor radius for six subsolar MPB crossings (SZA $<$ 30º). In all cases the normal is well defined ($\lambda_2 /\lambda_3 > 9$) and points mainly along the $\hat x$ axis. A case for the MPB thickness being of the order of the ion inertial length as well as the Larmor radius could be made for all crossings.

\begin{table}[h]
\setlength\tabcolsep{5pt}
\caption{In the successive columns, the following data of the six MPB crossings are displayed: date, time, minimum and maximum MPB thickness, ion inertial length, Larmor radius, volume current density, Lorentz force per unit volume, work done by the Lorentz force per volume unit, kinetic energy of solar wind ions upstream from the MPB.}
\centering
\begin{tabular}{l c l c l c l c l c l c l c l c}
\hline
Date & $\frac{t_2+t_3}{2}$ & $h_{23}$ & $h_{14}$ & $c/\omega_{pi}$ & $r_g$ & $|\mathbf{j}_v|$ & $|\mathbf{F}|$ & $W$ & $E_k$ \\
& \tiny{(UTC)} & \tiny{(km)} & \tiny{(km)} & \tiny{(km)} & \tiny{(km)} & \tiny{(nA/m$^2$)} & \tiny{(N/m$^3$)} & \tiny{(J/m$^3$)} & \tiny{(J/m$^3$)} \\
\hline
\hline
2015-Oct-10 & 12:41:58 & 39 & 97 & 159 & 203 & 403 & 4.37$\times 10^{-14}$ & 1.7 $\times 10^{-9}$ & 7.1 $\times 10^{-12}$ \footnotemark \\
2015-Oct-12 & 19:19:09 & 19 & 73 & 133 & 62 & 255 & 3.38$\times 10^{-14}$ & 6.4 $\times 10^{-10}$ & 9.6 $\times 10^{-12}$\\
2016-Mar-16 & 18:14:40 & 82 & 174 & 98 & 68 & 282 & 1.20$\times 10^{-14}$ & 9.8 $\times 10^{-10}$ & 1.1 $\times 10^{-11}$\\
2016-Mar-31 & 13:04:25 &39 & 122 & 101 & 76 & 401 & 1.34$\times 10^{-14}$ & 5.2 $\times 10^{-10}$ & 4.4 $\times 10^{-11}$\\
2016-Apr-05 & 05:16:22 & 44 & 175 & 130 & 168 & 363 & 0.92$\times 10^{-14}$&  4.0 $\times 10^{-10}$ & 2.7 $\times 10^{-12}$\\
2017-Nov-24 & 12:15:06 & 115 & 447 & 120 & 46 & 92 & 0.24$\times 10^{-14}$ & 2.8 $\times 10^{-10}$ & 8.9 $\times 10^{-12}$\\
\hline
\end{tabular}
\label{t:thickness}
\end{table}
\footnotetext[1]{This value was obtained using SWICS data as there is no SWICA data available for the selected crossing.}

\subsection{Current Density and the Lorentz Force at the MPB}
We estimated the current density along the boundary from Amp\`ere's Law in a discontinuity, assuming that the MPB is a planar surface of negligible thickness. 
If $ \hat n $ is the surface normal and $ \mathbf{B} _u $, $ \mathbf{B} _d $ are the magnetic field measurements upstream -in the magnetosheath- and downstream -in the MPR- respectively, the surface current density $ \mathbf{j} _s $ will be given by

\begin{equation}
\textbf{j}_s = \frac{1}{\mu_0}\hat n \times (\textbf{B}_u - \textbf{B}_d)
\end{equation} 

We calculated $ \mathbf{B} _u $ by taking the average value of \textbf{B} between 18:12:06 and 18:13:00 UTC and $ \mathbf{B} _d $ between 18:14:51 and 18:15:45 UTC;
these intervals are shaded in Fig. \ref{fig:zoom}. The values thus obtained were $ \mathbf{B} _u = (4.58, 19.24, -6.2) ~\textrm{nT} $ and $ \mathbf{B} _d = (18.64, 31.9, -28.59) ~\textrm{nT} $. 
The intervals were selected because they were outside the MPB but without large variations in the magnetic field, in order to be representative of what happens at the boundary.

The surface current density obtained based on the MVA normal yields $ \mathbf{j} _s ^ \textrm{MVA} = (-2.859,	-19.188, -12.640) ~\textrm{mA/m} $ and its magnitude  $ | \mathbf{j} _s ^ \textrm{MVA} | = 23.2 ~\textrm{mA/m} $.
Whereas, when the fit normal is used,  $ \mathbf{j} _s ^\textrm{fit} = (3.976, -20.983, -9.365) ~\textrm{mA/m}$, $ | \mathbf{j} _s ^\textrm{fit} | = 23.3 ~\textrm{mA/m} $. 

Under the previous approximations one can think of $ \mathbf{j} _s $ being constant throughout the MPB. In that case, a volume current density can simply be estimated as $ \mathbf{j}_s / h = \mathbf{j} _v $; we considered for this the minimum thickness $ h _{23}$ yielded by both the MVA and the fit. 
Using the MVA normal we obtained $ \mathbf{j} _v ^\textrm{MVA} = (-35, -233, -154) ~\textrm{nA/m}^2$ and its magnitude $ | \mathbf{j} _v ^\textrm{MVA} | = 282 ~\textrm{nA/m}^2$.
On the other hand, using the fit normal we obtained $ \mathbf{j} _v ^{fit} = (48, -255, -113) ~\textrm{nA/m}^2 $ with modulus $ | \mathbf{j} _v ^{fit} | = 284 ~\textrm{nA/m}^2 $.

The values of $ j_s $ and $ j_v $ obtained with both methods are consistent not only between them but with the values we obtained for different MPB crossings (shown in table \ref{t:thickness}) and those given by \citet{bertucci2005} for an MPB crossing with SZA = 63º from MGS data where they obtained $ | \mathbf{j}_s | = 6.5 ~\textrm{mA/m} $, $ | \mathbf{j}_v | = 81 ~\textrm{nA/m}^2 $.

Next, we calculate the Lorentz force per unit volume as $ \mathbf{F} = \mathbf{j}_v \times \mathbf{B} $.
From $ \mathbf{j}_v ^\textrm{MVA} $ we obtained the force $ \mathbf{F} ^\textrm{MVA} = (10.0, -3.3, 2.3) \times 10^{-15} ~\textrm{N/m}^3 $ and from $ \mathbf{j} _v^\textrm{fit} $ the force $ \mathbf{F} ^\textrm{fit} = (9.4, -0.7, 5.6) \times10^{-15} ~\textrm{N/m} ^3 $.

The work done by the Lorentz force along the MPB normal is $W = F_L h_{23} = 9.8 \times 10^{-10}$ J/m$^3$. This value is greater than the average kinetic energy of the solar wind protons in the magnetosheath upstream from the MPB (shaded in Fig. \ref{fig:zoom}), $E_k = \frac{1}{2} m_p v_n ^2 n = 1.1 \times 10 ^{-11}$ J/m$^3$, by almost two orders of magnitude; $v_n$ is the mean proton velocity in the direction of the MPB normal. Calculating the average kinetic energy of the solar wind protons before the shock (between 17:50 and 17:55 UTC), we find that it is $E_k = 4 \times 10^{-10}$ J/m$^3$, roughly half the work done by the Lorentz force. 

The Lorentz force is associated with the Hall term $ \mathbf{E} _H = \frac{1}{en} \mathbf{j} \times \mathbf{B} $ in the generalized Ohm's Law (eq. \ref{eq:ohm}). The force (and therefore, the Hall electric field) points mainly along the $ + \hat x $ axis, opposing the movement of the solar wind ions, which travel in $ - \hat x $, and accelerating the planetary ions.
The Hall electric field calculated from the values obtained through the MVA is $ \mathbf{E} _H ^ \textrm{MVA} = (26.37,-8.44,6.85) ~\textrm{mV/m} $ while the field calculated from the values obtained from the fit is $ \mathbf{E} _H ^ \textrm{fit} = (24.42,-1.85, 14.52) ~\textrm{mV/m} $.

\section{Discussion and Conclusions}
\label{section:discussion}

In this work we report on the microscopic properties of the Lorentz force associated with the current layer detected at the Martian MPB in the subsolar region from high-resolution data. The current is detected from the change in the tangential component of the magnetic field at the MPB. The intensities of the surface current density for the six analysed crossings range from 10.7 to 39.2 mA/m. This represents a factor two increase with respect to the values derived from MGS data by \citet{bertucci2005} closer to the terminator (6.5 mA/m at SZA = 63º). Although the sample is too small to deduce any general trend with SZA, the higher $\mathbf{j}_s$ values in the subsolar sector would be consistent with a stronger pileup \citep{dubinin2011} and/or a narrower MPB around the subsolar sector. The volume current density ranges from 92 nA/m$^2$ to 400 nA/m$^2$, up to 20 times greater than the values obtained by \citet{Ramstad2020}. Nonetheless, this discrepancy is to be expected as our study is centered on the fine structure of the MPB whereas theirs does not resolve structures smaller than 339 km. It too must be noted that as our selection consisted in crossings with a sharp increase in the magnetic field it may be biased towards greater values of \textbf{j}. 

A recent study by \citet{Haaland2020} shows a decrease in the Earth magnetopause current density with increasing SZA such that the current is two times stronger in the subsolar point than in the terminator.

Another key point is the thickness of the MPB. 
We find a strong variability in our estimates (from  18 km to 450 km) which is likely a result of the MPB moving with respect to the planet at speeds comparable to the spacecraft velocity during the crossing \citep{bertucci2005}. Unfortunately this effect cannot be corrected due to the nature of single spacecraft observations. Nevertheless, most cases display thicknesses that are loosely compatible with both the magnetosheath solar wind proton inertial length and with their gyro-radius (see Table \ref{t:thickness}). If the MPB thickness is somehow determined by  $c/\omega_{pi}$, a two-fluid MHD description should be able to theoretically capture this feature. On the other hand, if the thickness is determined by the Larmor radius, kinetic effects would need to be considered. The fact that these two length scales are not too dissimilar, makes it more difficult to discriminate between these two scenarios. A similar discussion takes place with the Earth magnetopause, as reported by \citet{Haaland2020} using MMS data for a large number of crossings. 

The magnetic pressure term in the Lorentz force is roughly inversely proportional to the MPB thickness while the magnetic tension is inversely proportional to the curvature radius of the magnetic field lines. As the MPB thickness is of the order of the hundred kilometers, while the typical radius of curvature of the draped magnetic field in the subsolar region is roughly 4000 km \citep{Vignes}, the first term will be at least one order of magnitude greater than the second. In the induced magnetotail, however, the magnetic tension dominates \citep{dubinin1993}.
 
In the six subsolar passes, the Lorentz force points in a direction not far from $\hat{x}$ (i.e. sunward) and its strength varies between 2.4$\times 10^{-15}$ N/m$^3$ and 4.37$\times 10^{-14}$ N/m$^3$. These values are one or two orders of magnitude stronger than the magnetic pressure gradients obtained by \citet{halekas_forces}. However, they report that their Lorentz force estimations might be underestimated as their values were averaged over large spatial intervals.
The work of the Lorentz force per unit volume is of the same order as the upstream mean kinetic density per unit volume in the solar wind while being at least an order higher than the mean kinetic density per unit volume upstream from the MPB. This strongly suggests that these ions can indeed be stopped by magnetic forces at the MPB in the subsolar sector.

A net force in the sunward direction contributes to the deceleration of the solar wind ions near the MPB while pushing the solar wind electrons inwards into the MPR. This would favor a decoupling between the solar wind protons and electrons (due to the Hall effect) as they struggle to enter the induced magnetosphere, while the solar wind electrons push the IMF through the MPB thus contributing to the magnetic barrier buildup \citep{dubinin2011}. In such a scenario the IMF would be frozen in to the electron plasma, not the ion plasma; quantifying this from direct measurements is a a major challenge even for multi-satelites missions such as MMS \citep{Lundin2005}.
In the meantime, quasi-neutrality across the MPB would be ensured by planetary ions which would be accelerated upwards by the sunward force. Some of these planetary ions would be able even to get out of the MPR although once in the magnetosheath they could be reaccelerated either by the electron pressure gradient (back into the MPR) or by the convective electric field into the plume \citep{dong2015}.

In summary, our results are consistent with a thickness for the martian MPB of the order of an ion inertial length. However, we cannot rule out the possibility that the MPB thickness is determined by the convective Larmor radius of solar wind protons, since: (1) these two length scales are not too dissimilar and, (2) we are bound by the limitations of single spacecraft observations. 

%
%
%
%
%
%
%
%

\acknowledgments
All data used are publicly available on the NASA Planetary Data System (https://pds.nasa.gov), under Search Data, MAVEN Mission, Planetary Plasma Interactions Node. 
The authors would like to thank the LIA-MAGNETO, CNRS-CONICET collaboration. G.B. is fellow of CONICET and C.B., L.M., D.O.G. are researchers of CONICET.
The authors acknowledge financial support from the Agencia de Promoción Científica y Tecnológica (Argentina) through grants PICT 1707/2015 and 1103/2018.


%
\bibliography{exampl.bib}

\begin{thebibliography}{}

\bibitem [\protect \citeauthoryear {%
Acu{\~n}a%
\ \protect \BOthers {.}}{%
Acu{\~n}a%
\ \protect \BOthers {.}}{%
{\protect \APACyear {1998}}%
}]{%
acuna}
\APACinsertmetastar {%
acuna}%
\begin{APACrefauthors}%
Acu{\~n}a, M\BPBI H.%
, Connerney, J\BPBI E\BPBI P.%
, Wasilewski, P.%
, Lin, R\BPBI P.%
, Anderson, K\BPBI A.%
, Carlson, C\BPBI W.%
\BDBL {}Ness, N\BPBI F.%
\end{APACrefauthors}%
\unskip\
\newblock
\APACrefYearMonthDay{1998}{}{}.
\newblock
{\BBOQ}\APACrefatitle {Magnetic Field and Plasma Observations at Mars: Initial
  Results of the Mars Global Surveyor Mission} {Magnetic field and plasma
  observations at mars: Initial results of the mars global surveyor
  mission}.{\BBCQ}
\newblock
\APACjournalVolNumPages{Science}{279}{5357}{1676--1680}.
\PrintBackRefs{\CurrentBib}

\bibitem [\protect \citeauthoryear {%
Anderson~Jr.%
}{%
Anderson~Jr.%
}{%
{\protect \APACyear {1974}}%
}]{%
anderson1974}
\APACinsertmetastar {%
anderson1974}%
\begin{APACrefauthors}%
Anderson~Jr., D\BPBI E.%
\end{APACrefauthors}%
\unskip\
\newblock
\APACrefYearMonthDay{1974}{}{}.
\newblock
{\BBOQ}\APACrefatitle {Mariner 6, 7, and 9 Ultraviolet Spectrometer Experiment:
  Analysis of hydrogen Lyman alpha data} {Mariner 6, 7, and 9 ultraviolet
  spectrometer experiment: Analysis of hydrogen lyman alpha data}.{\BBCQ}
\newblock
\APACjournalVolNumPages{Journal of Geophysical Research
  (1896-1977)}{79}{10}{1513-1518}.
\PrintBackRefs{\CurrentBib}

\bibitem [\protect \citeauthoryear {%
Anderson~Jr.%
\ \BBA {} Hord%
}{%
Anderson~Jr.%
\ \BBA {} Hord%
}{%
{\protect \APACyear {1971}}%
}]{%
anderson1971}
\APACinsertmetastar {%
anderson1971}%
\begin{APACrefauthors}%
Anderson~Jr., D\BPBI E.%
\BCBT {}\ \BBA {} Hord, C\BPBI W.%
\end{APACrefauthors}%
\unskip\
\newblock
\APACrefYearMonthDay{1971}{}{}.
\newblock
{\BBOQ}\APACrefatitle {Mariner 6 and 7 Ultraviolet Spectrometer Experiment:
  Analysis of hydrogen Lyman-alpha data} {Mariner 6 and 7 ultraviolet
  spectrometer experiment: Analysis of hydrogen lyman-alpha data}.{\BBCQ}
\newblock
\APACjournalVolNumPages{Journal of Geophysical Research
  (1896-1977)}{76}{28}{6666-6673}.
\PrintBackRefs{\CurrentBib}

\bibitem [\protect \citeauthoryear {%
Bertucci%
\ \protect \BOthers {.}}{%
Bertucci%
\ \protect \BOthers {.}}{%
{\protect \APACyear {2011}}%
}]{%
bertucci2011}
\APACinsertmetastar {%
bertucci2011}%
\begin{APACrefauthors}%
Bertucci, C.%
, Duru, F.%
, Edberg, N.%
, Fraenz, M.%
, Martinecz, C.%
, Szego, K.%
\BCBL {}\ \BBA {} Vaisberg, O.%
\end{APACrefauthors}%
\unskip\
\newblock
\APACrefYearMonthDay{2011}{}{}.
\newblock
{\BBOQ}\APACrefatitle {The Induced Magnetospheres of Mars, Venus, and Titan}
  {The induced magnetospheres of mars, venus, and titan}.{\BBCQ}
\newblock
\APACjournalVolNumPages{Space Science Reviews}{162}{1}{113--171}.
\PrintBackRefs{\CurrentBib}

\bibitem [\protect \citeauthoryear {%
Bertucci%
, Mazelle%
, Acuña%
, Russell%
\BCBL {}\ \BBA {} Slavin%
}{%
Bertucci%
\ \protect \BOthers {.}}{%
{\protect \APACyear {2005}}%
}]{%
bertucci2005}
\APACinsertmetastar {%
bertucci2005}%
\begin{APACrefauthors}%
Bertucci, C.%
, Mazelle, C.%
, Acuña, M\BPBI H.%
, Russell, C\BPBI T.%
\BCBL {}\ \BBA {} Slavin, J\BPBI A.%
\end{APACrefauthors}%
\unskip\
\newblock
\APACrefYearMonthDay{2005}{}{}.
\newblock
{\BBOQ}\APACrefatitle {Structure of the magnetic pileup boundary at Mars and
  Venus} {Structure of the magnetic pileup boundary at mars and venus}.{\BBCQ}
\newblock
\APACjournalVolNumPages{Journal of Geophysical Research: Space
  Physics}{110}{A1}{}.
\PrintBackRefs{\CurrentBib}

\bibitem [\protect \citeauthoryear {%
Bertucci%
\ \protect \BOthers {.}}{%
Bertucci%
\ \protect \BOthers {.}}{%
{\protect \APACyear {2003}}%
}]{%
bertucci2003a}
\APACinsertmetastar {%
bertucci2003a}%
\begin{APACrefauthors}%
Bertucci, C.%
, Mazelle, C.%
, Crider, D\BPBI H.%
, Vignes, D.%
, Acuña, M\BPBI H.%
, Mitchell, D\BPBI L.%
\BDBL {}Winterhalter, D.%
\end{APACrefauthors}%
\unskip\
\newblock
\APACrefYearMonthDay{2003}{}{}.
\newblock
{\BBOQ}\APACrefatitle {Magnetic field draping enhancement at the Martian
  magnetic pileup boundary from Mars global surveyor observations} {Magnetic
  field draping enhancement at the martian magnetic pileup boundary from mars
  global surveyor observations}.{\BBCQ}
\newblock
\APACjournalVolNumPages{Geophysical Research Letters}{30}{2}{}.
\PrintBackRefs{\CurrentBib}

\bibitem [\protect \citeauthoryear {%
Breus%
\ \protect \BOthers {.}}{%
Breus%
\ \protect \BOthers {.}}{%
{\protect \APACyear {1991}}%
}]{%
Breus1991}
\APACinsertmetastar {%
Breus1991}%
\begin{APACrefauthors}%
Breus, T\BPBI K.%
, Krymskii, A\BPBI M.%
, Lundin, R.%
, Dubinin, E\BPBI M.%
, Luhmann, J\BPBI G.%
, Yeroshenko, Y\BPBI G.%
\BDBL {}Styashkin, V\BPBI A.%
\end{APACrefauthors}%
\unskip\
\newblock
\APACrefYearMonthDay{1991}{}{}.
\newblock
{\BBOQ}\APACrefatitle {The solar wind interaction with Mars: Consideration of
  Phobos 2 mission observations of an ion composition boundary on the dayside}
  {The solar wind interaction with mars: Consideration of phobos 2 mission
  observations of an ion composition boundary on the dayside}.{\BBCQ}
\newblock
\APACjournalVolNumPages{Journal of Geophysical Research: Space
  Physics}{96}{A7}{11165-11174}.
\PrintBackRefs{\CurrentBib}

\bibitem [\protect \citeauthoryear {%
Cain%
, Ferguson%
\BCBL {}\ \BBA {} Mozzoni%
}{%
Cain%
\ \protect \BOthers {.}}{%
{\protect \APACyear {2003}}%
}]{%
Cain2003}
\APACinsertmetastar {%
Cain2003}%
\begin{APACrefauthors}%
Cain, J\BPBI C.%
, Ferguson, B\BPBI B.%
\BCBL {}\ \BBA {} Mozzoni, D.%
\end{APACrefauthors}%
\unskip\
\newblock
\APACrefYearMonthDay{2003}{}{}.
\newblock
{\BBOQ}\APACrefatitle {An n = 90 internal potential function of the Martian
  crustal magnetic field} {An n = 90 internal potential function of the martian
  crustal magnetic field}.{\BBCQ}
\newblock
\APACjournalVolNumPages{Journal of Geophysical Research: Planets}{108}{E2}{}.
\PrintBackRefs{\CurrentBib}

\bibitem [\protect \citeauthoryear {%
J.~Connerney%
\ \protect \BOthers {.}}{%
J.~Connerney%
\ \protect \BOthers {.}}{%
{\protect \APACyear {2015}}%
}]{%
MAG}
\APACinsertmetastar {%
MAG}%
\begin{APACrefauthors}%
Connerney, J.%
, Espley, J.%
, Lawton, P.%
, Murphy, S.%
, Odom, J.%
, Oliversen, R.%
\BCBL {}\ \BBA {} Sheppard, D.%
\end{APACrefauthors}%
\unskip\
\newblock
\APACrefYearMonthDay{2015}{}{}.
\newblock
{\BBOQ}\APACrefatitle {The MAVEN Magnetic Field Investigation} {The maven
  magnetic field investigation}.{\BBCQ}
\newblock
\APACjournalVolNumPages{}{195}{}{257-291}.
\PrintBackRefs{\CurrentBib}

\bibitem [\protect \citeauthoryear {%
J\BPBI E\BPBI P.~Connerney%
\ \protect \BOthers {.}}{%
J\BPBI E\BPBI P.~Connerney%
\ \protect \BOthers {.}}{%
{\protect \APACyear {2001}}%
}]{%
connerney2001}
\APACinsertmetastar {%
connerney2001}%
\begin{APACrefauthors}%
Connerney, J\BPBI E\BPBI P.%
, Acuña, M\BPBI H.%
, Wasilewski, P\BPBI J.%
, Kletetschka, G.%
, Ness, N\BPBI F.%
, Rème, H.%
\BDBL {}Mitchell, D\BPBI L.%
\end{APACrefauthors}%
\unskip\
\newblock
\APACrefYearMonthDay{2001}{}{}.
\newblock
{\BBOQ}\APACrefatitle {The global magnetic field of Mars and implications for
  crustal evolution} {The global magnetic field of mars and implications for
  crustal evolution}.{\BBCQ}
\newblock
\APACjournalVolNumPages{Geophysical Research Letters}{28}{21}{4015-4018}.
\PrintBackRefs{\CurrentBib}

\bibitem [\protect \citeauthoryear {%
Crider%
\ \protect \BOthers {.}}{%
Crider%
\ \protect \BOthers {.}}{%
{\protect \APACyear {2000}}%
}]{%
crider2000}
\APACinsertmetastar {%
crider2000}%
\begin{APACrefauthors}%
Crider, D.%
, Cloutier, P.%
, Law, C.%
, Walker, P.%
, Chen, Y.%
, Acuña, M.%
\BDBL {}Ness, N.%
\end{APACrefauthors}%
\unskip\
\newblock
\APACrefYearMonthDay{2000}{}{}.
\newblock
{\BBOQ}\APACrefatitle {Evidence of electron impact ionization in the magnetic
  pileup boundary of Mars} {Evidence of electron impact ionization in the
  magnetic pileup boundary of mars}.{\BBCQ}
\newblock
\APACjournalVolNumPages{Geophysical Research Letters}{27}{1}{45-48}.
\PrintBackRefs{\CurrentBib}

\bibitem [\protect \citeauthoryear {%
Dong%
\ \protect \BOthers {.}}{%
Dong%
\ \protect \BOthers {.}}{%
{\protect \APACyear {2015}}%
}]{%
dong2015}
\APACinsertmetastar {%
dong2015}%
\begin{APACrefauthors}%
Dong, Y.%
, Fang, X.%
, Brain, D\BPBI A.%
, McFadden, J\BPBI P.%
, Halekas, J\BPBI S.%
, Connerney, J\BPBI E.%
\BDBL {}Jakosky, B\BPBI M.%
\end{APACrefauthors}%
\unskip\
\newblock
\APACrefYearMonthDay{2015}{}{}.
\newblock
{\BBOQ}\APACrefatitle {Strong plume fluxes at Mars observed by MAVEN: An
  important planetary ion escape channel} {Strong plume fluxes at mars observed
  by maven: An important planetary ion escape channel}.{\BBCQ}
\newblock
\APACjournalVolNumPages{Geophysical Research Letters}{42}{21}{8942-8950}.
\PrintBackRefs{\CurrentBib}

\bibitem [\protect \citeauthoryear {%
Dubinin%
\ \protect \BOthers {.}}{%
Dubinin%
\ \protect \BOthers {.}}{%
{\protect \APACyear {2011}}%
}]{%
dubinin2011}
\APACinsertmetastar {%
dubinin2011}%
\begin{APACrefauthors}%
Dubinin, E.%
, Fraenz, M.%
, Fedorov, A.%
, Lundin, R.%
, Edberg, N.%
, Duru, F.%
\BCBL {}\ \BBA {} Vaisberg, O.%
\end{APACrefauthors}%
\unskip\
\newblock
\APACrefYearMonthDay{2011}{}{}.
\newblock
{\BBOQ}\APACrefatitle {Ion Energization and Escape on Mars and Venus} {Ion
  energization and escape on mars and venus}.{\BBCQ}
\newblock
\APACjournalVolNumPages{Space Science Reviews}{162}{}{173–211}.
\PrintBackRefs{\CurrentBib}

\bibitem [\protect \citeauthoryear {%
Dubinin%
, Lundin%
, Norberg%
\BCBL {}\ \BBA {} Pissarenko%
}{%
Dubinin%
\ \protect \BOthers {.}}{%
{\protect \APACyear {1993}}%
}]{%
dubinin1993}
\APACinsertmetastar {%
dubinin1993}%
\begin{APACrefauthors}%
Dubinin, E.%
, Lundin, R.%
, Norberg, O.%
\BCBL {}\ \BBA {} Pissarenko, N.%
\end{APACrefauthors}%
\unskip\
\newblock
\APACrefYearMonthDay{1993}{}{}.
\newblock
{\BBOQ}\APACrefatitle {Ion acceleration in the Martian tail: Phobos
  observations} {Ion acceleration in the martian tail: Phobos
  observations}.{\BBCQ}
\newblock
\APACjournalVolNumPages{Journal of Geophysical Research: Space
  Physics}{98}{A3}{3991-3997}.
\PrintBackRefs{\CurrentBib}

\bibitem [\protect \citeauthoryear {%
Dubinin%
\ \protect \BOthers {.}}{%
Dubinin%
\ \protect \BOthers {.}}{%
{\protect \APACyear {2008}}%
}]{%
dubinin2008}
\APACinsertmetastar {%
dubinin2008}%
\begin{APACrefauthors}%
Dubinin, E.%
, Modolo, R.%
, Fraenz, M.%
, Woch, J.%
, Duru, F.%
, Akalin, F.%
\BDBL {}Picardi, G.%
\end{APACrefauthors}%
\unskip\
\newblock
\APACrefYearMonthDay{2008}{}{}.
\newblock
{\BBOQ}\APACrefatitle {Structure and dynamics of the solar wind/ionosphere
  interface on Mars: MEX-ASPERA-3 and MEX-MARSIS observations} {Structure and
  dynamics of the solar wind/ionosphere interface on mars: Mex-aspera-3 and
  mex-marsis observations}.{\BBCQ}
\newblock
\APACjournalVolNumPages{Geophysical Research Letters}{35}{11}{}.
\PrintBackRefs{\CurrentBib}

\bibitem [\protect \citeauthoryear {%
Edberg%
, Lester%
, Cowley%
\BCBL {}\ \BBA {} Eriksson%
}{%
Edberg%
\ \protect \BOthers {.}}{%
{\protect \APACyear {2008}}%
}]{%
edberg2008}
\APACinsertmetastar {%
edberg2008}%
\begin{APACrefauthors}%
Edberg, N\BPBI J\BPBI T.%
, Lester, M.%
, Cowley, S\BPBI W\BPBI H.%
\BCBL {}\ \BBA {} Eriksson, A\BPBI I.%
\end{APACrefauthors}%
\unskip\
\newblock
\APACrefYearMonthDay{2008}{}{}.
\newblock
{\BBOQ}\APACrefatitle {Statistical analysis of the location of the Martian
  magnetic pileup boundary and bow shock and the influence of crustal magnetic
  fields} {Statistical analysis of the location of the martian magnetic pileup
  boundary and bow shock and the influence of crustal magnetic fields}.{\BBCQ}
\newblock
\APACjournalVolNumPages{Journal of Geophysical Research: Space
  Physics}{113}{A8}{}.
\PrintBackRefs{\CurrentBib}

\bibitem [\protect \citeauthoryear {%
Haaland%
\ \protect \BOthers {.}}{%
Haaland%
\ \protect \BOthers {.}}{%
{\protect \APACyear {2020}}%
}]{%
Haaland2020}
\APACinsertmetastar {%
Haaland2020}%
\begin{APACrefauthors}%
Haaland, S.%
, Paschmann, G.%
, Øieroset, M.%
, Phan, T.%
, Hasegawa, H.%
, Fuselier, S\BPBI A.%
\BDBL {}Burch, J.%
\end{APACrefauthors}%
\unskip\
\newblock
\APACrefYearMonthDay{2020}{}{}.
\newblock
{\BBOQ}\APACrefatitle {Characteristics of the Flank Magnetopause: MMS Results}
  {Characteristics of the flank magnetopause: Mms results}.{\BBCQ}
\newblock
\APACjournalVolNumPages{Journal of Geophysical Research: Space
  Physics}{125}{3}{e2019JA027623}.
\PrintBackRefs{\CurrentBib}

\bibitem [\protect \citeauthoryear {%
Halekas%
\ \protect \BOthers {.}}{%
Halekas%
\ \protect \BOthers {.}}{%
{\protect \APACyear {2017}}%
}]{%
halekas_forces}
\APACinsertmetastar {%
halekas_forces}%
\begin{APACrefauthors}%
Halekas, J\BPBI S.%
, Brain, D\BPBI A.%
, Luhmann, J\BPBI G.%
, DiBraccio, G\BPBI A.%
, Ruhunusiri, S.%
, Harada, Y.%
\BDBL {}Jakosky, B\BPBI M.%
\end{APACrefauthors}%
\unskip\
\newblock
\APACrefYearMonthDay{2017}{}{}.
\newblock
{\BBOQ}\APACrefatitle {Flows, Fields, and Forces in the Mars-Solar Wind
  Interaction} {Flows, fields, and forces in the mars-solar wind
  interaction}.{\BBCQ}
\newblock
\APACjournalVolNumPages{Journal of Geophysical Research: Space
  Physics}{122}{11}{11,320-11,341}.
\PrintBackRefs{\CurrentBib}

\bibitem [\protect \citeauthoryear {%
Halekas%
\ \protect \BOthers {.}}{%
Halekas%
\ \protect \BOthers {.}}{%
{\protect \APACyear {2018}}%
}]{%
halekas2018-ICB}
\APACinsertmetastar {%
halekas2018-ICB}%
\begin{APACrefauthors}%
Halekas, J\BPBI S.%
, McFadden, J\BPBI P.%
, Brain, D\BPBI A.%
, Luhmann, J\BPBI G.%
, DiBraccio, G\BPBI A.%
, Connerney, J\BPBI E\BPBI P.%
\BDBL {}Jakosky, B\BPBI M.%
\end{APACrefauthors}%
\unskip\
\newblock
\APACrefYearMonthDay{2018}{}{}.
\newblock
{\BBOQ}\APACrefatitle {Structure and Variability of the Martian Ion Composition
  Boundary Layer} {Structure and variability of the martian ion composition
  boundary layer}.{\BBCQ}
\newblock
\APACjournalVolNumPages{Journal of Geophysical Research: Space
  Physics}{123}{10}{8439-8458}.
\newblock
\begin{APACrefURL}
  \url{https://agupubs.onlinelibrary.wiley.com/doi/abs/10.1029/2018JA025866}
  \end{APACrefURL}
\newblock
\begin{APACrefDOI} \doi{10.1029/2018JA025866} \end{APACrefDOI}
\PrintBackRefs{\CurrentBib}

\bibitem [\protect \citeauthoryear {%
Halekas%
\ \protect \BOthers {.}}{%
Halekas%
\ \protect \BOthers {.}}{%
{\protect \APACyear {2015}}%
}]{%
SWIA}
\APACinsertmetastar {%
SWIA}%
\begin{APACrefauthors}%
Halekas, J\BPBI S.%
, Taylor, E\BPBI R.%
, Dalton, G.%
, Johnson, G.%
, Curtis, D\BPBI W.%
, McFadden, J\BPBI P.%
\BDBL {}Jakosky, B\BPBI M.%
\end{APACrefauthors}%
\unskip\
\newblock
\APACrefYearMonthDay{2015}{}{}.
\newblock
{\BBOQ}\APACrefatitle {The Solar Wind Ion Analyzer for MAVEN} {The solar wind
  ion analyzer for maven}.{\BBCQ}
\newblock
\APACjournalVolNumPages{Space Science Reviews}{195}{1}{125--151}.
\PrintBackRefs{\CurrentBib}

\bibitem [\protect \citeauthoryear {%
Harnett%
\ \BBA {} Winglee%
}{%
Harnett%
\ \BBA {} Winglee%
}{%
{\protect \APACyear {2007}}%
}]{%
Harnett2007}
\APACinsertmetastar {%
Harnett2007}%
\begin{APACrefauthors}%
Harnett, E\BPBI M.%
\BCBT {}\ \BBA {} Winglee, R\BPBI M.%
\end{APACrefauthors}%
\unskip\
\newblock
\APACrefYearMonthDay{2007}{}{}.
\newblock
{\BBOQ}\APACrefatitle {High-resolution multifluid simulations of the plasma
  environment near the Martian magnetic anomalies} {High-resolution multifluid
  simulations of the plasma environment near the martian magnetic
  anomalies}.{\BBCQ}
\newblock
\APACjournalVolNumPages{Journal of Geophysical Research: Space
  Physics}{112}{A5}{}.
\newblock
\begin{APACrefURL}
  \url{https://agupubs.onlinelibrary.wiley.com/doi/abs/10.1029/2006JA012001}
  \end{APACrefURL}
\newblock
\begin{APACrefDOI} \doi{10.1029/2006JA012001} \end{APACrefDOI}
\PrintBackRefs{\CurrentBib}

\bibitem [\protect \citeauthoryear {%
Holmberg%
\ \protect \BOthers {.}}{%
Holmberg%
\ \protect \BOthers {.}}{%
{\protect \APACyear {2019}}%
}]{%
Holmberg2019}
\APACinsertmetastar {%
Holmberg2019}%
\begin{APACrefauthors}%
Holmberg, M\BPBI K\BPBI G.%
, André, N.%
, Garnier, P.%
, Modolo, R.%
, Andersson, L.%
, Halekas, J.%
\BDBL {}Mitchell, D\BPBI L.%
\end{APACrefauthors}%
\unskip\
\newblock
\APACrefYearMonthDay{2019}{}{}.
\newblock
{\BBOQ}\APACrefatitle {MAVEN and MEX Multi-instrument Study of the Dayside of
  the Martian Induced Magnetospheric Structure Revealed by Pressure Analyses}
  {Maven and mex multi-instrument study of the dayside of the martian induced
  magnetospheric structure revealed by pressure analyses}.{\BBCQ}
\newblock
\APACjournalVolNumPages{Journal of Geophysical Research: Space
  Physics}{124}{11}{8564-8589}.
\PrintBackRefs{\CurrentBib}

\bibitem [\protect \citeauthoryear {%
{Jakosky}%
\ \protect \BOthers {.}}{%
{Jakosky}%
\ \protect \BOthers {.}}{%
{\protect \APACyear {2015}}%
}]{%
jakosky_MAVEN}
\APACinsertmetastar {%
jakosky_MAVEN}%
\begin{APACrefauthors}%
{Jakosky}, B\BPBI M.%
, {Lin}, R\BPBI P.%
, {Grebowsky}, J\BPBI M.%
, {Luhmann}, J\BPBI G.%
, {Mitchell}, D\BPBI F.%
, {Beutelschies}, G.%
\BDBL {}{Zurek}, R.%
\end{APACrefauthors}%
\unskip\
\newblock
\APACrefYearMonthDay{2015}{{\APACmonth{12}}}{}.
\newblock
{\BBOQ}\APACrefatitle {{The Mars Atmosphere and Volatile Evolution ( MAVEN)
  Mission}} {{The Mars Atmosphere and Volatile Evolution ( MAVEN)
  Mission}}.{\BBCQ}
\newblock
\APACjournalVolNumPages{Space Science Reviews}{195}{1-4}{3-48}.
\newblock
\begin{APACrefDOI} \doi{10.1007/s11214-015-0139-x} \end{APACrefDOI}
\PrintBackRefs{\CurrentBib}

\bibitem [\protect \citeauthoryear {%
Knetter%
, Neubauer%
, Horbury%
\BCBL {}\ \BBA {} Balogh%
}{%
Knetter%
\ \protect \BOthers {.}}{%
{\protect \APACyear {2004}}%
}]{%
knetter}
\APACinsertmetastar {%
knetter}%
\begin{APACrefauthors}%
Knetter, T.%
, Neubauer, F\BPBI M.%
, Horbury, T.%
\BCBL {}\ \BBA {} Balogh, A.%
\end{APACrefauthors}%
\unskip\
\newblock
\APACrefYearMonthDay{2004}{}{}.
\newblock
{\BBOQ}\APACrefatitle {Four-point discontinuity observations using Cluster
  magnetic field data: A statistical survey} {Four-point discontinuity
  observations using cluster magnetic field data: A statistical survey}.{\BBCQ}
\newblock
\APACjournalVolNumPages{Journal of Geophysical Research: Space
  Physics}{109}{A6}{}.
\PrintBackRefs{\CurrentBib}

\bibitem [\protect \citeauthoryear {%
{Lundin}%
, {Yamauchi}%
, {Sauvaud}%
\BCBL {}\ \BBA {} {Balogh}%
}{%
{Lundin}%
\ \protect \BOthers {.}}{%
{\protect \APACyear {2005}}%
}]{%
Lundin2005}
\APACinsertmetastar {%
Lundin2005}%
\begin{APACrefauthors}%
{Lundin}, R.%
, {Yamauchi}, M.%
, {Sauvaud}, J\BPBI A.%
\BCBL {}\ \BBA {} {Balogh}, A.%
\end{APACrefauthors}%
\unskip\
\newblock
\APACrefYearMonthDay{2005}{{\APACmonth{10}}}{}.
\newblock
{\BBOQ}\APACrefatitle {Magnetospheric plasma boundaries: a test of the
  frozen-in magnetic field theorem} {Magnetospheric plasma boundaries: a test
  of the frozen-in magnetic field theorem}.{\BBCQ}
\newblock
\APACjournalVolNumPages{Annales Geophysicae}{23}{7}{2565-2578}.
\newblock
\begin{APACrefDOI} \doi{10.5194/angeo-23-2565-2005} \end{APACrefDOI}
\PrintBackRefs{\CurrentBib}

\bibitem [\protect \citeauthoryear {%
Mahaffy%
\ \protect \BOthers {.}}{%
Mahaffy%
\ \protect \BOthers {.}}{%
{\protect \APACyear {2015}}%
}]{%
mahaffy}
\APACinsertmetastar {%
mahaffy}%
\begin{APACrefauthors}%
Mahaffy, P\BPBI R.%
, Benna, M.%
, Elrod, M.%
, Yelle, R\BPBI V.%
, Bougher, S\BPBI W.%
, Stone, S\BPBI W.%
\BCBL {}\ \BBA {} Jakosky, B\BPBI M.%
\end{APACrefauthors}%
\unskip\
\newblock
\APACrefYearMonthDay{2015}{}{}.
\newblock
{\BBOQ}\APACrefatitle {Structure and composition of the neutral upper
  atmosphere of Mars from the MAVEN NGIMS investigation} {Structure and
  composition of the neutral upper atmosphere of mars from the maven ngims
  investigation}.{\BBCQ}
\newblock
\APACjournalVolNumPages{Geophysical Research Letters}{42}{21}{8951-8957}.
\PrintBackRefs{\CurrentBib}

\bibitem [\protect \citeauthoryear {%
Matsunaga%
\ \protect \BOthers {.}}{%
Matsunaga%
\ \protect \BOthers {.}}{%
{\protect \APACyear {2017}}%
}]{%
matsunaga2017}
\APACinsertmetastar {%
matsunaga2017}%
\begin{APACrefauthors}%
Matsunaga, K.%
, Seki, K.%
, Brain, D\BPBI A.%
, Hara, T.%
, Masunaga, K.%
, Mcfadden, J\BPBI P.%
\BDBL {}Jakosky, B\BPBI M.%
\end{APACrefauthors}%
\unskip\
\newblock
\APACrefYearMonthDay{2017}{}{}.
\newblock
{\BBOQ}\APACrefatitle {Statistical Study of Relations Between the Induced
  Magnetosphere, Ion Composition, and Pressure Balance Boundaries Around Mars
  Based On MAVEN Observations} {Statistical study of relations between the
  induced magnetosphere, ion composition, and pressure balance boundaries
  around mars based on maven observations}.{\BBCQ}
\newblock
\APACjournalVolNumPages{Journal of Geophysical Research: Space
  Physics}{122}{9}{9723-9737}.
\PrintBackRefs{\CurrentBib}

\bibitem [\protect \citeauthoryear {%
Mitchell%
\ \protect \BOthers {.}}{%
Mitchell%
\ \protect \BOthers {.}}{%
{\protect \APACyear {2016}}%
}]{%
SWEA}
\APACinsertmetastar {%
SWEA}%
\begin{APACrefauthors}%
Mitchell, D.%
, Mazelle, C.%
, Sauvaud, J.%
, Thocaven, J.%
, Rouzaud, J.%
, Fedorov, A.%
\BDBL {}Jakosky, B.%
\end{APACrefauthors}%
\unskip\
\newblock
\APACrefYearMonthDay{2016}{}{}.
\newblock
{\BBOQ}\APACrefatitle {The MAVEN Solar Wind Electron Analyzer} {The maven solar
  wind electron analyzer}.{\BBCQ}
\newblock
\APACjournalVolNumPages{}{200}{}{495-528}.
\PrintBackRefs{\CurrentBib}

\bibitem [\protect \citeauthoryear {%
Moses%
, Coroniti%
\BCBL {}\ \BBA {} Scarf%
}{%
Moses%
\ \protect \BOthers {.}}{%
{\protect \APACyear {1988}}%
}]{%
moses1988}
\APACinsertmetastar {%
moses1988}%
\begin{APACrefauthors}%
Moses, S\BPBI L.%
, Coroniti, F\BPBI V.%
\BCBL {}\ \BBA {} Scarf, F\BPBI L.%
\end{APACrefauthors}%
\unskip\
\newblock
\APACrefYearMonthDay{1988}{}{}.
\newblock
{\BBOQ}\APACrefatitle {Expectations for the microphysics of the Mars-solar wind
  interaction} {Expectations for the microphysics of the mars-solar wind
  interaction}.{\BBCQ}
\newblock
\APACjournalVolNumPages{Geophysical Research Letters}{15}{5}{429-432}.
\PrintBackRefs{\CurrentBib}

\bibitem [\protect \citeauthoryear {%
Neubauer%
}{%
Neubauer%
}{%
{\protect \APACyear {1987}}%
}]{%
Neubauer1987}
\APACinsertmetastar {%
Neubauer1987}%
\begin{APACrefauthors}%
Neubauer, F\BPBI M.%
\end{APACrefauthors}%
\unskip\
\newblock
\APACrefYearMonthDay{1987}{}{}.
\newblock
{\BBOQ}\APACrefatitle {Giotto magnetic-field results on the boundaries of the
  pile-up region and the magnetic cavity} {Giotto magnetic-field results on the
  boundaries of the pile-up region and the magnetic cavity}.{\BBCQ}
\newblock
\APACjournalVolNumPages{Astronomy and astrophysics}{187}{}{73-79}.
\PrintBackRefs{\CurrentBib}

\bibitem [\protect \citeauthoryear {%
Ramstad%
\ \protect \BOthers {.}}{%
Ramstad%
\ \protect \BOthers {.}}{%
{\protect \APACyear {2020}}%
}]{%
Ramstad2020}
\APACinsertmetastar {%
Ramstad2020}%
\begin{APACrefauthors}%
Ramstad, R.%
, Brain, D.%
, Dong, Y.%
, Espley, J.%
, Halekas, J.%
\BCBL {}\ \BBA {} Jakosky, B.%
\end{APACrefauthors}%
\unskip\
\newblock
\APACrefYearMonthDay{2020}{}{}.
\newblock
{\BBOQ}\APACrefatitle {The global current systems of the Martian induced
  magnetosphere} {The global current systems of the martian induced
  magnetosphere}.{\BBCQ}
\newblock
\APACjournalVolNumPages{Nature Astronomy}{}{}{}.
\PrintBackRefs{\CurrentBib}

\bibitem [\protect \citeauthoryear {%
Ruhunusiri%
\ \protect \BOthers {.}}{%
Ruhunusiri%
\ \protect \BOthers {.}}{%
{\protect \APACyear {2017}}%
}]{%
Ruhunusiri2017}
\APACinsertmetastar {%
Ruhunusiri2017}%
\begin{APACrefauthors}%
Ruhunusiri, S.%
, Halekas, J\BPBI S.%
, Espley, J\BPBI R.%
, Mazelle, C.%
, Brain, D.%
, Harada, Y.%
\BDBL {}Howes, G\BPBI G.%
\end{APACrefauthors}%
\unskip\
\newblock
\APACrefYearMonthDay{2017}{}{}.
\newblock
{\BBOQ}\APACrefatitle {Characterization of turbulence in the Mars plasma
  environment with MAVEN observations} {Characterization of turbulence in the
  mars plasma environment with maven observations}.{\BBCQ}
\newblock
\APACjournalVolNumPages{Journal of Geophysical Research: Space
  Physics}{122}{1}{656-674}.
\PrintBackRefs{\CurrentBib}

\bibitem [\protect \citeauthoryear {%
Sauer%
, Bogdanov%
\BCBL {}\ \BBA {} Baumgärtel%
}{%
Sauer%
\ \protect \BOthers {.}}{%
{\protect \APACyear {1994}}%
}]{%
Sauer1994}
\APACinsertmetastar {%
Sauer1994}%
\begin{APACrefauthors}%
Sauer, K.%
, Bogdanov, A.%
\BCBL {}\ \BBA {} Baumgärtel, K.%
\end{APACrefauthors}%
\unskip\
\newblock
\APACrefYearMonthDay{1994}{}{}.
\newblock
{\BBOQ}\APACrefatitle {Evidence of an ion composition boundary (protonopause)
  in bi-ion fluid simulations of solar wind mass loading} {Evidence of an ion
  composition boundary (protonopause) in bi-ion fluid simulations of solar wind
  mass loading}.{\BBCQ}
\newblock
\APACjournalVolNumPages{Geophysical Research Letters}{21}{20}{2255-2258}.
\PrintBackRefs{\CurrentBib}

\bibitem [\protect \citeauthoryear {%
Sonnerup%
\ \BBA {} Scheible%
}{%
Sonnerup%
\ \BBA {} Scheible%
}{%
{\protect \APACyear {1998}}%
}]{%
sonnerup_scheible}
\APACinsertmetastar {%
sonnerup_scheible}%
\begin{APACrefauthors}%
Sonnerup, B\BPBI U.%
\BCBT {}\ \BBA {} Scheible, M.%
\end{APACrefauthors}%
\unskip\
\newblock
\APACrefYearMonthDay{1998}{}{}.
\newblock
{\BBOQ}\APACrefatitle {Minimum and maximum variance analysis} {Minimum and
  maximum variance analysis}.{\BBCQ}
\newblock
\BIn{} \APACrefbtitle {Analysis Methods for Multi-Spacecraft Data} {Analysis
  methods for multi-spacecraft data}\ (\BPG~185-220).
\newblock
\APACaddressPublisher{}{ESA Publications Division}.
\PrintBackRefs{\CurrentBib}

\bibitem [\protect \citeauthoryear {%
Szego%
\ \protect \BOthers {.}}{%
Szego%
\ \protect \BOthers {.}}{%
{\protect \APACyear {2000}}%
}]{%
szego}
\APACinsertmetastar {%
szego}%
\begin{APACrefauthors}%
Szego, K.%
, Tsurutani, B.%
, Bogdanov, A.%
, Bingham, R.%
, Haerendel, G.%
, Brinca, A.%
\BDBL {}Fischer, C.%
\end{APACrefauthors}%
\unskip\
\newblock
\APACrefYearMonthDay{2000}{}{}.
\newblock
{\BBOQ}\APACrefatitle {Physics of Mass Loaded Plasmas} {Physics of mass loaded
  plasmas}.{\BBCQ}
\newblock
\APACjournalVolNumPages{Space Science Reviews}{94}{}{429-671}.
\newblock
\begin{APACrefDOI} \doi{10.1023/A:1026568530975} \end{APACrefDOI}
\PrintBackRefs{\CurrentBib}

\bibitem [\protect \citeauthoryear {%
Trotignon%
, Mazelle%
, Bertucci%
\BCBL {}\ \BBA {} Acuña%
}{%
Trotignon%
\ \protect \BOthers {.}}{%
{\protect \APACyear {2006}}%
}]{%
trotignon2006}
\APACinsertmetastar {%
trotignon2006}%
\begin{APACrefauthors}%
Trotignon, J.%
, Mazelle, C.%
, Bertucci, C.%
\BCBL {}\ \BBA {} Acuña, M.%
\end{APACrefauthors}%
\unskip\
\newblock
\APACrefYearMonthDay{2006}{}{}.
\newblock
{\BBOQ}\APACrefatitle {Martian shock and magnetic pile-up boundary positions
  and shapes determined from the Phobos 2 and Mars Global Surveyor data sets}
  {Martian shock and magnetic pile-up boundary positions and shapes determined
  from the phobos 2 and mars global surveyor data sets}.{\BBCQ}
\newblock
\APACjournalVolNumPages{}{54}{}{357-369}.
\newblock
\begin{APACrefDOI} \doi{10.1016/j.pss.2006.01.003} \end{APACrefDOI}
\PrintBackRefs{\CurrentBib}

\bibitem [\protect \citeauthoryear {%
Vignes%
\ \protect \BOthers {.}}{%
Vignes%
\ \protect \BOthers {.}}{%
{\protect \APACyear {2000}}%
}]{%
Vignes}
\APACinsertmetastar {%
Vignes}%
\begin{APACrefauthors}%
Vignes, D.%
, Mazelle, C.%
, Rme, H.%
, Acuña, M\BPBI H.%
, Connerney, J\BPBI E\BPBI P.%
, Lin, R\BPBI P.%
\BDBL {}Ness, N\BPBI F.%
\end{APACrefauthors}%
\unskip\
\newblock
\APACrefYearMonthDay{2000}{}{}.
\newblock
{\BBOQ}\APACrefatitle {The solar wind interaction with Mars: Locations and
  shapes of the bow shock and the magnetic pile-up boundary from the
  observations of the MAG/ER Experiment onboard Mars Global Surveyor} {The
  solar wind interaction with mars: Locations and shapes of the bow shock and
  the magnetic pile-up boundary from the observations of the mag/er experiment
  onboard mars global surveyor}.{\BBCQ}
\newblock
\APACjournalVolNumPages{Geophysical Research Letters}{27}{1}{49-52}.
\PrintBackRefs{\CurrentBib}

\end{thebibliography}

%


\end{document}